\documentclass{article}
\usepackage{amssymb}
\usepackage{amsmath}
\usepackage{amsfonts}
\usepackage{authblk}
\usepackage{enumerate}

\begin{document}
\title{Complete integrability of geodesic motion in Sasaki-Einstein 
toric $Y^{p,q}$ spaces}

\author{Elena Mirela Babalic\thanks{mbabalic@theory.nipne.ro}~}
\author{Mihai Visinescu\thanks{mvisin@theory.nipne.ro}}

\affil{Department of Theoretical Physics,

National Institute for Physics and Nuclear Engineering,

Magurele, P.O.Box M.G.-6, Romania}

\date{\today }
\maketitle

\begin{abstract}

We construct explicitly the constants of motion for geodesics in the $5$-dimensional 
Sasaki-Einstein spaces $Y^{p,q}$. To carry out this task we use the knowledge of
the complete set of Killing vectors and Killing-Yano tensors on these spaces.
In spite of the fact that we generate a multitude of constants of motion, only 
five of them are functionally independent implying the complete integrability of
geodesic flow on $Y^{p,q}$ spaces. In the particular case of the homogeneous 
Sasaki-Einstein  manifold $T^{1,1}$ the integrals of motion have simpler forms and 
the relations between them are described in detail.

\end{abstract}

~


~


~

\section{Introduction}

In the last years, there has been a considerable interest in finding explicit 
examples of Sasaki-Einstein spaces which provide supersymmetric background relevant
to the AdS/CFT conjecture \cite{JM}. Among them, we mention the construction of 
inhomogeneous Sasaki-Einstein metrics \cite{GMSW1,GMSW2}. In particular, an 
interesting class is represented by the toric contact structures on $S^2 \times S^3$
denoted $Y^{p,q}$, where $q < p$ are positive integers.

The purpose of this paper is the explicit construction of the constants of motion for 
geodesics in $Y^{p,q}$ spaces. This goal is achieved using the complete set of Killing 
vectors and Killing-Yano tensors on these $5$-dimensional toric Sasaki-Einstein spaces.
The importance of Killing forms comes from the fact that it is possible to associate
with them St\"{a}ckel-Killing tensors of rank $2$. In this way we have conserved 
quantities for geodesic motion associated with manifest and also hidden symmetries.

In spite of the fact that we are able to construct many integrals of motion, it turns 
out that only a part of them are functionally independent. We get that in general
on $Y^{p,q}$ spaces there are only five independent constants of motion, providing the 
complete integrability of geodesic flow, but not superintegrability.

In order to get a better understanding of the relations between various constants of 
motion, we resort to the particular case of the homogeneous Sasaki-Einstein manifold
$T^{1,1}$. The constants of geodesic motion on $T^{1,1}$ have simpler forms and the 
relations among them is explicitly described.

The paper is organized as follows. In the next Section we introduce the main concepts
and technical tools we use in the rest of the paper. We give the necessary preliminaries 
regarding Killing tensors and toric Sasaki-Einstein manifolds. In Section 3 we 
investigate the constants of motion for geodesics on $Y^{p,q}$ spaces proving the
complete integrability. In Section 4 we resume the construction of integrals of 
motion on the homogeneous Sasaki-Einstein space $T^{1,1}$. The expressions of 
constants of motion are simpler allowing us to write explicitly the relations among them.
Finally, our conclusions are presented within the last Section.

\section{Preliminaries}
\subsection{St\"{a}ckel-Killing and Killing-Yano tensors}

Killing vector fields represent a basic object of differential geometry connected 
with the infinitesimal isometries. Let $(M,g)$ be an $n$-dimensional Riemannian 
manifold with the metric $g$ and let $\nabla$ be its Levi-Civita connection. 
The vector field $K_\mu$ preserves the metric $g$ if it satisfies Killing's equation
\begin{equation}
\nabla_{(\mu} K_{\nu)} =0 \,.
\end{equation}
Here and elsewhere a round bracket denotes a symmetrization over the indices within.

The geodesic quadratic Hamiltonian which parametrizes the inverse metric is:
\begin{equation}\label{Ham}
H = \frac12 g^{\mu\nu} P_\mu P_\nu \,,
\end{equation}
where $P_\mu$ are canonical momenta conjugate to the coordinates $x^\mu$,
$P_\mu = g_{\mu\nu}\dot{x}^\nu$ with overdot denoting proper time derivative.

In the presence of a Killing vector, the system of a free particle admits a 
conserved quantity
\begin{equation}
K= K_\mu \dot{x}^\mu \,,
\end{equation}
which commutes with the Hamiltonian \eqref{Ham} in the sense of Poisson brackets:
\begin{equation}\label{PB}
\{K,H\} = 0\,.
\end{equation}

A natural generalization of Killing vector fields is represented by 
St\"{a}ckel-Killing tensors. A St\"{a}ckel-Killing tensor is a covariant symmetric 
tensor field satisfying the generalized Killing equation
\begin{equation}\label{SK}
\nabla_{(\lambda}K_{\mu_1,\dots ,\mu_r)} = 0\,.
\end{equation}
St\"{a}ckel-Killing tensors may be identified with first integrals of the
Hamiltonian geodesic flow, which are homogeneous polynomials in the momenta.
Again, using \eqref{PB} and the generalized Killing equation \eqref{SK} we 
get that the quantities
\begin{equation}\label{SKcons}
K_{SK} = K_{\mu_1 \dots\mu_r}\dot{x}^{\mu_1}\cdots \dot{x}^{\mu_r}\,,
\end{equation}
are constants of geodesic motion.

The next most simple objects that can be studied in connection with the 
symmetries of a manifold after the Killing vectors and St\"{a}ckel-Killing 
tensors are the Killing forms or Killing-Yano tensors. These are differential 
$r$-forms satisfying the equation
\begin{equation}\label{KY}
\nabla_{(\mu} f_{\nu_1)\nu_2\dots \nu_r} =0\,. 
\end{equation}
It can be easily verified that along every geodesic in $M$,
$f_{\mu \nu_2\dots \nu_r} \dot{x}^\mu$ is parallel.

Let us note that there is an important connection between these two 
generalizations of the Killing vectors. To wit, given two Killing-Yano tensors 
$f^{(1)}_{\phantom{(1)}\nu_1\nu_2\dots \nu_r}$ and 
$f^{(2)}_{\phantom{(1)}\nu_1\nu_2\dots \nu_r}$ there is
a St\"{a}ckel-Killing tensor of rank $2$
\begin{equation}\label{KYY}
K^{(1,2)}_{\mu\nu} = f^{(1)}_{\phantom{(1)}\mu\lambda_2 \dots \lambda_r}
f_{\phantom{(2)}\nu}^{(2)\phantom{\nu}\lambda_2 \dots \lambda_r} +
f^{(2)}_{\phantom{(2)}\mu\lambda_2 \dots \lambda_r}
f_{\phantom{(1)}\nu}^{(1)\phantom{\nu}\lambda_2 \dots \lambda_r}
\,.
\end{equation}

This fact offers a method to generate higher order integrals of motion by 
identifying the complete set of Killing-Yano tensors. We shall use this
technique in the case of the toric Sasaki-Einstein $Y^{p,q}$ spaces for which
the complete set of Killing-Yano tensors is known \cite{Vis,S-V-V}.

The multitude of Killing-Yano tensors allows us to construct the conserved
quantities for the geodesic motion and investigate the integrability of the 
$Y^{p,q}$ geometry.

Let us recall that in classical mechanics a Hamiltonian system with Hamiltonian 
$H$ \eqref{Ham} and integrals of motion $K_j$ is called {\it completely integrable}
(or Liouville integrable) if it allows $n$ integrals of motion $H, K_1, 
\dots, K_{n-1}$ which are well-defined functions on the phase space, in involution
\begin{equation}
\{H, K_j\} = 0\,,~~ \{K_j, K_k\} = 0\,,\ ~~j,k = 1,\cdots , n-1\,,
\end{equation}
and functionally independent. A system is {\it superintegrable} if it is 
completely integrable and allows further functionally independent integrals of motion.

\subsection{Toric Sasaki-Einstein spaces}

A $(2n-1)$-dimensional manifold $M$ is a \emph{contact manifold} if
there exists a $1$-form $\eta$ (called a \emph{contact $1$-form}) on
$M$ such that:
\begin{equation}
\eta \wedge (d \eta)^{n-1} \neq 0\,.
\end{equation}
For every choice of contact $1$-form $\eta$ there exists a unique
vector field $K_{\eta}$, called the {\it Reeb vector field}, which satisfies:
\begin{equation}
\eta (K_{\eta}) = 1 \quad \text{and} \quad K_{\eta} \lrcorner d\eta = 0\,,
\end{equation}
where $\lrcorner$ is the operator dual to the wedge product.

A contact Riemannian manifold $(M, g_M)$ is Sasakian if its metric 
cone $C(M)$  
\begin{equation}
C(M) \cong \mathbb{R}_+ \times M\,,~~ g_{C(M)} = dr^2 + r^2\,g_{M} \,, 
\end{equation}
is K\"{a}hler \cite{BG} with K\"{a}hler form \cite{MS,MSY}
\begin{equation}\label{omega}
\omega = \frac 12 d(r^2 \eta) =r dr \wedge \eta + \frac 12 r^2 d\eta \,.
\end{equation}
Here $r \in (0,\infty)$ may be considered as a coordinate on the
positive real line $\mathbb{R}_+$.
The Sasakian manifold $(M, g_{M})$ is naturally isometrically embedded into
the metric cone via the inclusion
\begin{equation}
M = \{ r=1 \} = \{ 1 \} \times M \subset C(M)\,.
\end{equation}

In the case of Sasaki-Einstein manifolds there is a well known fact 
that the following statements are equivalent \cite{FOW}:
\begin{enumerate}[(i)]
\item $(M,g_M)$ is Sasaki-Einstein with $\mathop{\rm Ric} g_M = 2(n-1) 
g_M$;
\item The K\"{a}hler cone $(C(M),g_{C(M)})$ is Ricci-flat
($\mathop{\rm Ric} g_{C(M)} = 0$), i.e. a Calabi-Yau manifold;
\item The transverse K\"{a}hler structure to the Reeb foliation
$\mathcal{F}_{K_\eta}$ is K\"{a}hler-Einstein with 
$\mathop{\rm Ric}g^T = 2 n g^T$.
\end{enumerate}

The metric cone $C(M)$ is called toric if the standard $n$-torus 
$\mathbb{T}^n = \mathbb{R}^n/ 2 \pi \mathbb{Z}^n$ acts effectively on 
it, preserving the K\"{a}hler form $\omega$.

On a $(2n-1)$-dimensional Sasaki manifold with the contact
$1$-form $\eta$ there are the following Killing forms \cite{Semm}:
\begin{equation}\label{Psik}
\Psi_k = \eta \wedge (d\eta)^k \quad , \quad k = 0, 1, 
\cdots , n-1\,.
\end{equation}

Let us note that in the case of the Calabi-Yau cone the holonomy is $SU(n)$
and there are {\it two additional} parallel forms of degree $n$. In order
to write explicitly the additional Killing forms which correspond to these 
parallel forms, we shall express the
volume form of the metric cone in terms of the K\"{a}hler form
(\ref{omega})
\begin{equation}
d\mathcal{V}=\frac{1}{n!}\omega^n\,.
\end{equation}
Here $\omega^n$ is the wedge product of $\omega$ with itself $n$ times.
The volume of a K\"{a}hler manifold can also be written as
\cite{Semm,MVGV}
\begin{equation}
d\mathcal{V} = \frac{i^n}{2^n} (-1)^{n(n-1)/2} \Omega \wedge
\overline{\Omega}\,,
\end{equation}
where $\Omega$ is the holomorphic $(n,0)$ volume form
of $C(M)$. The additional (real) parallel forms are given by the real and
respectively imaginary part of the complex volume form.

Finally, to extract the corresponding additional Killing forms of the
Einstein-Sasaki spaces, we make use of the fact that for any $p$-form
$\psi$ on the space $M$ we can define an associated $p+1$-form $\psi^C$
on the cone $C(M)$ \cite{Semm}:
\begin{equation}\label{CPsi}
\psi^C := r^p d r \wedge \psi + \frac{r^{p+1}}{p+1} d \psi .
\end{equation}

\section{Complete integrability on $Y^{p,q}$ spaces}

Let us consider the explicit local metric of the 5-dimensional 
$Y(p,q)$ manifold given by the square of the line element \cite{GMSW1,GMSW2,BK,RS}
\begin{equation}\label{Ypq}
\begin{split}
ds^2_{ES} & = \frac{1-c\, y}{6}( d \theta^2 + \sin^2 \theta\, d \phi^2) 
+  \frac{1}{w(y)q(y)} dy^2 
+ \frac{q(y)}{9} ( d\psi - \cos \theta \, d \phi)^2 \nonumber\\
& \quad  + 
w(y)\left[ d\alpha + \frac{ac -2y+ c\, y^2}{6(a-y^2)}
(d\psi - \cos\theta \, d\phi)\right]^2\,,
\end{split}
\end{equation}
where
\begin{equation}
w(y)  = \frac{2(a-y^2)}{1-cy} \,, \quad 
q(y)  = \frac{a-3y^2 + 2c y^3}{a-y^2}\,.
\end{equation}
This metric is Einstein with $\mathop{\rm Ric} g = 4 g$ for all values of the
constants $a,c$. For $c=0$ the metric takes the 
local form of the standard homogeneous metric on $T^{1,1}$ 
\cite{M-S}. Otherwise the constant $c$ can be rescaled by a 
diffeomorphism and in what follows we assume $c=1$. For
\begin{equation}
0 < \alpha < 1~,
\end{equation}
we can take the range of the angular coordinates $(\theta, \phi, \psi)$
to be $0\leq \theta \leq 2\pi\,, 0\leq \phi \leq 2\pi\,, 0\leq \psi \leq
2\pi$. Choosing $0 < a < 1$,  the roots $y_i$  of the cubic equation
\begin{equation}
a - 3 y^2 + 2 y^3 = 0 \,
\end{equation}
are real, one negative $(y_1)$ and two positive $(y_2, y_3)$. If the
smallest of the positive roots is $y_2$, one can take the range of the
coordinate $y$ to be
\begin{equation}
y_1\leq y \leq y_2 \,.
\end{equation}
The Reeb vector is :\cite{M-S}
\begin{equation}\label{Reeb}
K = 3 \frac {\partial}{\partial \psi}
- \frac{1}{2}\frac {\partial}{\partial \alpha}\,,
\end{equation}
and the Sasakian $1$-form $\eta$ is \cite{M-S}:
\begin{equation}
\eta = -2y d\alpha + \frac{1-y}3 (d\psi - \cos\theta d\phi)\,.
\end{equation}
The conjugate momenta to the coordinates $(\theta,\phi, y, \alpha, \psi)$
are:
\begin{equation}\label{momenta}
\begin{split}
&P_{\theta} =
\frac{1-y}{6} \dot{\theta}\,,\\
&P_{\phi} + \cos\theta P_{\psi} = \frac{1-y}{6} \sin^2\theta \dot{\phi}\,,\\ 
&P_y = \frac{1}{6 p(y)} \dot{y}\,,\\
&P_{\alpha}=w(y) \left(\dot{\alpha} + f(y)  \left(\dot{\psi} - \cos\theta
\dot{\phi}\right)\right) \,,\\
&P_{\psi} = w(y) f(y) \dot{\alpha} +
\left[ \frac{q(y)}{9} + w(y) f^2(y)\right]\left(\dot{\psi} - \cos\theta
\dot{\phi}\right)\,,
\end{split}
\end{equation}
where 
\begin{equation}
\begin{split}
f(y)=&\frac{a-2y+y^2}{6(a-y^2)} \,,\\  
p(y)=& \frac{w(y) q(y)}{6} = \frac{a-3y^2+2y^3}{3(1-y)} \,.
\end{split}
\end{equation}
Using the momenta \eqref{momenta}, the conserved Hamiltonian \eqref{Ham}
becomes:
\begin{equation}
\begin{split}
H=&\frac12 \Biggl\{ 6 p(y) P_y^2 + \frac{6}{1-y}\bigl(P_\theta^2 + 
\frac{1}{\sin^2 \theta}(P_\phi  + \cos\theta P_\psi)^2\bigr) + 
\frac{1-y}{2(a-y^2)}P^2_\alpha\Biggr.\\
& \Biggl.+ \frac{9(a-y^2)}{a-3y^2 +2 y^3}\biggl(P_\psi -
\frac{a -2y +y^2}{6(a-y^2)} P_\alpha \biggr)^2\Biggr\}\\
=&\frac{1-y}{12} (\dot{\theta}^2+\sin^2\theta \dot{\phi}^2)+ 
\frac{\dot{y}^2}{12 p(y)}
+ \frac{q(y)}{18} (\dot{\psi} - \cos\theta \dot{\phi})^2 \\ 
&+ \frac{w(y)}{2}[\dot{\alpha}+f(y)(\dot{\psi}-\cos\theta \dot{\phi})]^2\,.
\end{split}
\end{equation}
From the isometry $SU(2) \times U(1) \times U(1)$ of the metric \eqref{Ypq}
we have that the momenta $P_\phi, P_\psi$ and $P_\alpha$ are conserved.
$P_\phi$ is the third component of the $SU(2)$ angular momentum and 
$P_\psi, P_\alpha$ are associated to the $U(1)$ factors. In addition,
the total $SU(2)$ angular momentum 
\begin{equation}
\vec{J}^{~2} =P_{\theta}^2 + \frac{1}{\sin^2\theta} \left(P_{\phi}+ 
\cos\theta P_{\psi}\right)^2 + P_{\psi}^2  \,
\end{equation}
is also conserved.

The next conserved quantities, quadratic in momenta, will be expressed
in terms of St\"{a}ckel-Killing tensors as in \eqref{SKcons}. The 
St\"{a}ckel-Killing tensors of rank two on $Y^{p,q}$ will be constructed 
from Killing-Yano tensors according to \eqref{KYY}. For this purpose we
shall use the Killing-Yano tensor $\Psi_1$ from \eqref{Psik} for $k=1$
and the additional parallel forms of degree $2$, associated with the real 
and imaginary parts of the holomorphic $(3,0)$ volume form
$\Omega$ of the cone $C(Y^{p,q})$.

The explicit form of the Killing-Yano tensor $\Psi_1$ is
\begin{equation}\label{Psi1}
\begin{split}
\Psi_1 & = (1-  y)^2\sin\theta \, d\theta \wedge d\phi \wedge 
d\psi - 6 dy \wedge d \alpha \wedge d \psi \\ 
& \quad + 6 \cos\theta \,
d\phi\wedge dy \wedge d\alpha - 6 (1-  y) y 
\sin\theta \, d\theta \wedge d\phi \wedge d\alpha\,.
\end{split}
\end{equation}

Let us call $\Psi$ the Killing form on $Y^{p,q}$ related to the complex holomorphic 
$(3,0)$ form on $C(Y^{p,q})$. The real and imaginary parts of $\Psi$ are 
\cite{Vis,S-V-V}:
\begin{equation}\label{RePsi}
\begin{split}
\Re \Psi &
= \sqrt{\frac{1 - y}{p(y)}}\\
&\quad \times \biggl( \cos\psi
\Bigl[d\theta \wedge dy  + 6 p(y) \sin\theta \, 
d\phi \wedge d\alpha +  p(y) \sin\theta \, d\phi \wedge d \psi \Bigr]  \biggr.\\
& \quad \biggl. -\sin\psi\Bigl[\sin\theta \, d\phi \wedge dy
- 6p(y) d\theta \wedge d\alpha -p(y) d\theta \wedge d\psi \Bigr. \biggr. \\
& \Bigl.\biggl.~~~~~~~~~~~~~+ p(y) \cos\theta \, d\theta \wedge d\phi \Bigr] \biggr)\,,
\end{split}
\end{equation}
\begin{equation}\label{ImPsi}
\begin{split}
\Im\Psi &
= \sqrt{\frac{1 - y}{p(y)}}\\
&\quad \times \biggl( \sin\psi
\Bigl[d\theta \wedge dy  + 6 p(y) \sin\theta \, 
d\phi \wedge d\alpha +  p(y) \sin\theta \, d\phi \wedge d \psi \Bigr]  \biggr.\\
& \quad \biggl. +\cos\psi\Bigl[\sin\theta \, d\phi \wedge dy
- 6p(y) d\theta \wedge d\alpha -p(y) d\theta \wedge d\psi \Bigr. \biggr. \\
& \Bigl.\biggl.~~~~~~~~~~~~~+ p(y) \cos\theta \, d\theta \wedge d\phi \Bigr] \biggr)\,.
\end{split}
\end{equation}

The first St\"{a}ckel-Killing tensor $(K1)_{\mu\nu}$ is constructed according to 
\eqref{KYY} using the real part of the Killing form $\Psi$:
\begin{equation}\label{K1RR}
(K1)_{\mu\nu} =(\Re\Psi)_{\mu\lambda} (\Re\Psi)_{\phantom{\lambda}\nu}^{\lambda}
\end{equation}
and has the non-vanishing components:
\begin{equation}\label{K1munu}
\begin{split}
K1_{\theta\theta} &=  6(1-y) \,,\\
K1_{\phi\phi} &=  \frac{3 +a -6y +2y^3 + (-3 +a +6y - 6y^2 + 2 y^3)\cos 2\theta}
{1-y}\,,\\
K1_{\phi\alpha}&= K1_{\alpha\phi} =- 12\frac{(a + (-3+2y)y^2)\cos\theta}{1-y}\,,\\
K1_{\phi\psi}&= K1_{\psi\phi} = -2\frac{(a + (-3+2y)y^2)\cos\theta}{1-y} \,,\\
K1_{yy} &= 18\frac{1-y}{a + (-3 +2y)y^2}\,,\\
K1_{\alpha\alpha} &= 72\frac{a + (-3+2y)y^2}{1-y}\,,\\
K1_{\alpha\psi} &= K1_{\psi\alpha} = 12\frac{a + (-3+2y)y^2}{1-y}\,,\\
K1_{\psi\psi} &= 2\frac{a + (-3+2y)y^2}{1-y}\,.
\end{split}
\end{equation}
The corresponding conserved quantity \eqref{SKcons} is
\begin{equation}\label{K1}
\begin{split}
K1 =& 6(1-y)\dot{\theta}\dot{\theta}  + \frac{3 +a -6y +2y^3 + 
(-3 +a +6y - 6y^2 + 2 y^3)\cos 2\theta}{1-y}\dot{\phi}\dot{\phi}\\
&-24\frac{(a + (-3+2y)y^2)\cos\theta}{1-y}\dot{\phi}\dot{\alpha}
- 4\frac{(a + (-3+2y)y^2)\cos\theta}{1-y}\dot{\phi}\dot{\psi}\\
&+ 18\frac{1-y}{a + (-3 +2y)y^2}\dot{y}\dot{y}
+72\frac{a + (-3+2y)y^2}{1-y}\dot{\alpha}\dot{\alpha}\\
&+24\frac{a + (-3+2y)y^2}{1-y}\dot{\alpha}\dot{\psi}
+2\frac{a + (-3+2y)y^2}{1-y}\dot{\psi}\dot{\psi}\,.
\end{split}
\end{equation}

The next St\"{a}ckel-Killing tensor  will be constructed from the imaginary part of 
$\Psi$:
\begin{equation}\label{K2II}
(K2)_{\mu\nu} =(\Im\Psi)_{\mu\lambda} (\Im\Psi)_{\phantom{\lambda}\nu}^{\lambda}\,,
\end{equation}
and we find that this tensor has the same components as $K1$ 
\eqref{K1munu}.

The mixed combination of $\Re \Psi$ and $\Im \Psi$ produces the St\"{a}ckel-Killing 
tensor
\begin{equation}\label{K3RI}
(K3)_{\mu\nu} =(\Re\Psi)_{\mu\lambda} (\Im\Psi)_{\phantom{\lambda}\nu}^{\lambda}
+ (\Im\Psi)_{\mu\lambda} (\Re\Psi)_{\phantom{\lambda}\nu}^{\lambda}\,,
\end{equation}
but it proves that all components of this tensor vanish.

Finally we construct the St\"{a}ckel-Killing tensor from the Killing form $\Psi_1$:
\begin{equation}\label{K4PP}
(K4)_{\mu\nu} =(\Psi_1)_{\mu\lambda\sigma} 
(\Psi_1)_{\phantom{\lambda\sigma}\nu}^{\lambda\sigma}\,.
\end{equation}
The non-vanishing components of this tensor are:\footnote{In \cite{RS} the components 
of this tensors are correct, but the expression of the conserved quantity contains 
some misprints. Consequently, the evaluation of the number of functionally 
independent set of integrals of motion is affected.}
\begin{equation}\label{K4munu}
\begin{split}
K4_{\theta\theta} &= 108(1-y) \,,\\
K4_{\phi\phi} &= 18\frac{7 +a -18y + 12 y^2 -2y^3 + (1 +a-6y+6y^2-2 y^3)\cos 2\theta}
{1-y}\,,\\
K4_{\phi\alpha}&= K4_{\alpha\phi} = - 216\frac{(a + (-4 +5y -2y^2)y)\cos\theta}{1-y}\,,\\
K4_{\phi\psi}&= K4_{\psi\phi} = - 36\frac{(a -(2-y)^2 (-1+2y))\cos\theta}{1-y} \,,\\
K4_{yy} &=  324\frac{1-y}{a + (-3 +2y)y^2}\,,\\
K4_{\alpha\alpha} &= 1296\frac{a + (1-2y)y^2}{1-y}\,,\\
K4_{\alpha\psi} &= K4_{\psi\alpha} = 216\frac{a + (-4+5y-2y^2)y}{1-y}\,,\\
K4_{\psi\psi} &= 36\frac{a -(2-y)^2 (-1+2y)}{1-y}\,.
\end{split}
\end{equation}
and the corresponding conserved quantity is:
\begin{equation}\label{K4}
\begin{split}
K4 =& 108(1-y)\dot{\theta}\dot{\theta}\\
&+18\frac{7 +a -18y + 12 y^2 -2y^3 + 
(1 +a-6y+6y^2-2 y^3)\cos 2\theta}{1-y}\dot{\phi}\dot{\phi}\\
&- 432\frac{(a + (-4 +5y -2y^2)y)\cos\theta}{1-y}\dot{\phi}\dot{\alpha}\\
&- 72\frac{(a -(2-y)^2 (-1+2y))\cos\theta}{1-y}\dot{\phi}\dot{\psi}\\
&+ 324\frac{1-y}{a + (-3 +2y)y^2}\dot{y}\dot{y}
+1296\frac{a + (1-2y)y^2}{1-y}\dot{\alpha}\dot{\alpha}\\
& +432\frac{a + (-4+5y-2y^2)y}{1-y}\dot{\alpha}\dot{\psi}
+36\frac{a -(2-y)^2 (-1+2y)}{1-y}\dot{\psi}\dot{\psi}\,.
\end{split}
\end{equation}

Having in mind that $K1=K2$ and $K3$ vanishes, we shall verify if the set 
$H, P_{\phi},P_{\psi}, P_{\alpha}, \vec{J}^{~2}, K1, K4$ constitutes a 
functionally independent set of constants of motion for the geodesics of
$Y^{p,q}$ constructing the Jacobian:
\begin{equation}
\mathcal{J} = \frac{\partial(H, P_{\phi},P_{\psi}, P_{\alpha},
\vec{J}^{~2}, K1, K4)}
{\partial(\theta,\phi, y, \alpha,\psi,
\dot{\theta}, \dot{\phi}, \dot{y}, \dot{\alpha}, \dot{\psi})}\,.
\end{equation}
Evaluating the rank of this Jacobian we find:
\begin{equation}
\text{Rank}~ \mathcal{J} = 5\,,
\end{equation}
which means that the system is completely integrable.
In spite of the presence of the St\"{a}ckel-Killing tensors $K1$ and $K4$
the system is not superintegrable, $K1$ and $K4$ being a combination of the 
first integrals $H, P_{\phi},P_{\psi}, P_{\alpha}, \vec{J}^{~2}$.

In the next Section we shall analyze the integrability for the space $T^{1,1}$.
In this case the formulas are not so intricate and the dependence of the 
corresponding St\"{a}ckel-Killing tensors $K1$ and $K4$ of the set
$H, P_{\phi},P_{\psi}, P_{\alpha}, \vec{J}^{~2}$ will be worked out explicitly.

\section{Complete integrability on $T^{1,1}$ space}

The homogeneous Sasaki-Einstein metric on $S^2 \times S^3$ is usually referred 
to as $T^{1,1}$. The $T^{1,1}$ space was considered as the first example of
toric Sasaki-Einstein/quiver duality \cite{KW}.

The isometries of $T^{1,1}$ form the group $SU(2) \times SU(2) \times U(1)$
and the metric of this space may be written down explicitly by utilizing the 
fact that it is a $U(1)$ bundle over $S^2 \times S^2$. Let us denote by
$(\theta_1,\phi_1)$ and $(\theta_2,\phi_2)$ the coordinates which parametrize 
the two sphere in a conventional way, and the angle $\psi \in [0, 4 \pi)$ to 
parametrize the $U(1)$ fiber. Using these parametrizations the $T^{1,1}$ metric
may be written as \cite{C-O,M-S}:
\begin{equation}
\begin{split}
ds^2(T^{1,1}) = & \frac16 (d \theta^2_1 + \sin^2 \theta_1 d \phi^2_1 +
d \theta^2_2 + \sin^2 \theta_2 d \phi^2_2) +\\
& \frac19 (d \psi + \cos \theta_1 d \phi_1 + \cos \theta_2 d \phi_2)^2
\,.
\end{split}
\end{equation}
The globally defined contact $1$-form $\eta$ is:
\begin{equation}\label{etaT}
\eta =\frac 13(d \psi +\cos \theta _1 d \phi _1+\cos \theta _2 d \phi _2)\,,
\end{equation}
and the Reeb vector field $K_\eta$ has the form:
\begin{equation}\label{ReebT}
K_\eta = 3 \frac{\partial}{\partial \psi} \,
\end{equation}
and it is easy to see that $\eta(K_\eta)=1$.

The conjugate momenta to the coordinates $(\theta_1,\theta_2,\phi_1,\phi_2,\psi)$
are:
\begin{equation}
\begin{split}
P_{\theta_1} &= \frac16 \dot{\theta}_1\,,\\
P_{\theta_2} &= \frac16 \dot{\theta}_2\,, \\
P_{\phi_1} &= \frac16 \sin^2\theta_1 \,\dot{\phi}_1 + 
\frac19 \cos^2\theta_1 \,\dot{\phi}_1
+ \frac19 \cos\theta_1 \,\dot{\psi} + 
\frac19 \cos\theta_1 \cos\theta_2\,\dot{\phi}_2\,,\\
P_{\phi_2} &= \frac16 \sin^2\theta_2 \,\dot{\phi}_2 + 
\frac19 \cos^2\theta_2 \,\dot{\phi}_2
+ \frac19 \cos\theta_2 \,\dot{\psi} + 
\frac19 \cos\theta_1 \cos\theta_2 \,\dot{\phi}_1\,,\\
P_{\psi} &= \frac19 \,\dot{\psi} + \frac19\cos\theta_1 \,\dot{\phi}_1 + 
\frac19\cos\theta_2 \,\dot{\phi}_2\,,\label{Ppsi}
\end{split}
\end{equation}
and the conserved Hamiltonian \eqref{Ham} takes the form:
\begin{equation}\label{H1}
\begin{split}
H=& 3 \left[ P^2_{\theta_1} +  P^2_{\theta_2} +
\frac{1}{\sin^2\theta_1}( P_{\phi_1} - \cos\theta_1 P_{\psi})^2 +
\frac{1}{\sin^2\theta_2}( P_{\phi_2} - \cos\theta_2 P_{\psi})^2 \right] \\
& + \frac92P^2_{\psi} \\
=& \frac{1}{12} (\dot{\theta}^2_1 + \sin^2 \theta_1 \dot{\phi}^2_1 +
\dot{\theta}^2_2 + \sin^2 \theta_2 \dot{\phi}^2_2) 
+ \frac{1}{18} (\dot{\psi} + \cos\theta_1\dot{\phi}_1 +\cos\theta_2\dot{\phi}_2)^2
\,.
\end{split}
\end{equation}

Taking into account the isometries of $T^{1,1}$, the momenta $P_{\phi_1},P_{\phi_2}$
and $P_{\psi}$ are conserved. On the other hand two total $SU(2)$ angular momenta 
are also conserved:
\begin{equation}
\begin{split}
\vec{J}_1^{~2} =& P_{\theta_1}^2 + \frac{1}{\sin^2\theta_1}( P_{\phi_1}
- \cos\theta_1 P_{\psi})^2  +P^2_{\psi}\\
=&\frac{1}{36}\biggl[\dot{\theta}^2_1  +\sin^2\theta_1 \dot{\phi}^2_1 \biggr]
+ \frac{1}{81}\biggl[\dot{\psi}^2 + \cos^2\theta_1 \dot{\phi}^2_1 + 
\cos^2\theta_2 \dot{\phi}^2_2 \biggr.\\
&\biggl. + 2\cos\theta_1\, \dot{\phi}_1 \dot{\psi} +
2 \cos\theta_2\dot{\phi}_2\dot{\psi} + 
2 \cos\theta_1 \cos\theta_2 \dot{\phi}_1\dot{\phi}_2\biggr]\,;\\
\vec{J}_2^{~2} =& P_{\theta_2}^2 + \frac{1}{\sin^2\theta_2}( P_{\phi_2}
- \cos\theta_2 P_{\psi})^2  +P^2_{\psi}\\
=&\frac{1}{36}\biggl[\dot{\theta}^2_2  +\sin^2\theta_2 \dot{\phi}^2_2 \biggr]
+ \frac{1}{81}\biggl[\dot{\psi}^2 + \cos^2\theta_1 \dot{\phi}^2_1 + 
\cos^2\theta_2 \dot{\phi}^2_2 \biggr.\\
&\biggl. + 2\cos\theta_1\, \dot{\phi}_1 \dot{\psi} +
2 \cos\theta_2\dot{\phi}_2\dot{\psi} + 
2 \cos\theta_1 \cos\theta_2 \dot{\phi}_1\dot{\phi}_2\biggr].
\end{split}
\end{equation}

In order to construct the St\"{a}ckel-Killing tensors on $T^{1,1}$ we shall 
consider its metric cone $C(T^{1,1})$. The Calabi-Yau cone over $T^{1,1}$
has the complex structure of the quadric singularity 
$\{ z_1^2 + z_2^2 + z_3^2 + z_4^2=0\} \subset \mathbb{C}^4$ minus the 
isolated singular point at the origin. From the holomorphic
$(3,0)$ volume form on the Calabi-Yau cone $C(T^{1,1})$ we extract the additional 
Killing form $\Psi$ on the $T^{1,1}$ making use of the prescription 
\eqref{CPsi}. After standard calculations we get the real and imaginary parts of 
$\Psi$ \cite{SVV-epl}:
\begin{equation}
\begin{split}
\Re \Psi =&\cos \psi \,d\theta _1\wedge d\theta _2+\sin \theta _2\sin \psi
\,d\theta_1\wedge d\phi _2 \\
& -\sin \theta _1\sin \psi \,d\theta _2\wedge d\phi _1 \\
&-\sin \theta _1\sin\theta _2\cos \psi \,d\phi _1\wedge d\phi _2\,,
\end{split}
\end{equation}
\begin{equation}
\begin{split}
\Im \Psi  =&\sin \psi \,d\theta _1\wedge d\theta _2-\sin \theta _2\cos \psi
\,d\theta _1\wedge d\phi _2 \\
& +\sin \theta _1\cos \psi \,d\theta _2\wedge d\phi _1 \\
&-\sin \theta _1\sin \theta _2\sin \psi \,d\phi _1\wedge d\phi _2\,.
\end{split}
\end{equation}
Using these Killing forms, the  St\"{a}ckel-Killing tensor \eqref{K1RR} becomes in this case:
\begin{equation}
(K1)_{\mu\nu} = 6\left(
\begin{array}{ccccc}
1&0&0&0&0\\
0&1&0&0&0\\
0&0&\sin^2\theta_1&0&0\\
0&0&0&\sin^2\theta_2&0\\
0&0&0&0&0
\end{array}
\right) \,.
\end{equation}

As in the case of $Y^{p,q}$ spaces, we observe that the St\"{a}ckel-Killing tensor
$K2$ \eqref{K2II} has the same components as $K1$ while $K3$ \eqref{K3RI} vanishes.

On the other hand, from the contact form $\eta$ \eqref{etaT} we evaluate the 
Killing-Yano tensor $\Psi_1$ \cite{SVV-epl}:
\begin{equation}
\begin{split}
\Psi _1=&\frac 19(\sin \theta _1 d\psi \wedge d\theta_1\wedge d\phi _1
+\sin \theta _2 d\psi \wedge d\theta _2\wedge d\phi _2  \\
&-\cos \theta _1\sin \theta _2 d\theta _2\wedge d\phi _1\wedge d\phi_2  \\
&+\cos \theta _2\sin \theta _1 d\theta _1\wedge d\phi _1\wedge d\phi _2)\,.
\end{split}
\end{equation}
The corresponding St\"{a}ckel-Killing tensor $K4$ \eqref{K4PP} is:
\begin{equation}
(K4)_{\mu\nu} = \frac43 \left(
\begin{array}{ccccc}
1 & 0 & 0 & 0 & 0 \\
0 &1& 0 & 0 & 0 \\
0&0&\frac13(3 + \cos^2\theta_1)&\frac43\cos\theta_1\cos\theta_2&\frac43\cos\theta_1\\
0&0&\frac43\cos\theta_1\cos\theta_2&\frac13(3 +\cos^2\theta_2)&\frac43\cos\theta_2\\
0&0&\frac43\cos\theta_1&\frac43\cos\theta_2&\frac43
\end{array}
\right) \,.
\end{equation}

Finally we investigate the integrability of the geodesic motion on $T^{1,1}$
and for this purpose we construct the Jacobian:
\begin{equation}
\mathcal{J} = \frac{\partial(H, P_{\phi_1}, P_{\phi_2}, P_{\psi}, 
\vec{J}_1^{~2}, \vec{J}_2^{~2}, K1, K4)}
{\partial(\theta_1,\theta_2,\phi_1, \phi_2,\psi,
\dot{\theta}_1, \dot{\theta}_2, \dot{\phi}_1, \dot{\phi}_2, \dot{\psi})}\,.
\end{equation}

As expected, we get that the rank of this Jacobian is $5$ implying the complete integrability of
the geodesic motion on $T^{1,1}$. Therefore not all aforesaid constants of motion are 
functionally independent. For example we can choose the subset $(H, P_{\phi_1}, 
P_{\phi_2}, P_{\psi}, \vec{J}_1^{~2})$ as functionally independent constants of motion.
It is quite simple to verify that the constants of motion $\vec{J}_2^{~2}, K1$ and $K4$
are combinations of the chosen subset of constants:
\begin{equation}
\frac16 K1 = 12 \mathcal{H} - \frac23(9 P_{\psi})^2\,,
\end{equation}
\begin{equation}
\frac34 K4 = 12 \mathcal{H} + \frac23(9 P_{\psi})^2\,,
\end{equation}
\begin{equation}
6 \vec{J}_2^{~2} = 2 H + 3 P_{\psi}^2 - 6 \vec{J}_1^{~2} \,.
\end{equation}

Therefore the study of integrability of geodesic motion in $T^{1,1}$ reconfirms
the results obtained in the previous section, making them more clear and convincing.

\section{Conclusions}

In this paper we presented the complete set of constants of motion for geodesics
in $Y^{p,q}$ spaces. In the particular case of $T^{1,1}$ space the formulas are
not so intricate and the discussion of the relations between various constants
of motion is greatly simplified.

The complete integrability of geodesic equations for the spaces considered is 
closely related to the property of the complete separation of variables in some 
field equations on these spaces. The functionally independent first integrals 
on the phase space are in involution and the system is completely integrable in 
the Liouville sense. 

It would be interesting to extend these investigations to other higher dimensional 
Sasaki-Einstein spaces relevant for predictions of the AdS/CFT correspondence 
\cite{M-S-Y}. The integrability of the geodesic flow on Sasaki-Einstein spaces
offer new perspectives in the investigation of supersymmetries and separability 
of Hamilton-Jacobi, Klein-Gordon and Dirac equations on these spaces.

\section*{Acknowledgments}

The authors are grateful to Vladimir Slesar and Gabriel-Eduard V\^{i}lcu
for their useful comments in the early stages of this work. 
M.V. was supported by CNCS-UEFISCDI, project number PN-II-ID-PCE-2011-3-0137 
while the work of E.M.B. was funded by  CNCS-UEFISCDI project number 
PN-II-ID-PCE-2011-3-0264 and by PN 09  37 01 02/2009.


\begin{thebibliography}{99}
%
\bibitem{JM}
J. M. Maldacena,
\textit{Adv. Theor. Math. Phys.} \textbf{2} (1998) 231.
%
\bibitem{GMSW1}
J. P. Gauntlett, D. Martelli, J. Sparks and D. Waldram,
\textit{Class. Quant. Grav.}  \textbf{21} (2004) 4335.
%
\bibitem{GMSW2}
J. P. Gauntlett, D. Martelli, J. Sparks and D. Waldram,
\textit{Adv. Theor. Math. Phys.}  \textbf{8} (2004) 711.
%
\bibitem{Vis}  M. Visinescu, 
\textit{Mod. Phys. Lett. A} \textbf{27} (2012) 1250217.
%
\bibitem{S-V-V}
V. Slesar, M. Visinescu and G. E. V\^ilcu,
\textit{Phys. Scripta} \textbf{89} (2014) 125205.
%
\bibitem{BG}
C. P. Boyer and K. Galicki, 
\textit{Sasakian geometry}, Oxford Mathematical Monographs, Oxford 
University Press, Oxford, 2008.
%
\bibitem{MS}
D. Martelli and  J. Sparks,
\textit{Phys. Lett. B} \textbf{621} (2005) 208.
%
\bibitem{MSY} 
D. Martelli, J. Sparks and S.-T. Yau, 
\textit{Comm. Math. Phys.} \textbf{280} (2008) 611.
%
\bibitem{FOW}
A. Futaki, H. Ono and G. Wang,
\textit{J. Diff. Geom.} {\bf 83} (2009) 585.
%
\bibitem{Semm}  
U. Semmelmann, 
\textit{Math. Z.} \textbf{245} (2003) 503.
%
\bibitem{MVGV}
M. Visinescu and  G. E. V\^{\i}lcu,
\textit{SIGMA} {\bf 8} (2012) 058. 
%
\bibitem{BK} 
S. Benvenuti and M. Kruczenski,
\textit{JHEP} \textbf{0610} (2006) 051. 
%
\bibitem{RS} 
E. Rub\'{i}n de Celis and O. P. Santill\'{a}n,
\textit{JHEP} \textbf{09} (2012) 032.
%
\bibitem{M-S}  
D. Martelli and J. Sparks, 
\textit{Comm. Math. Phys.} \textbf{262} (2006) 51.
%
\bibitem{KW}
I. R. Klebanov and E. Witten, 
\textit{Nucl. Phys.} \textbf{B536} (1998) 199.
%
\bibitem{C-O}
P. Candelas and X. de la Ossa, 
\textit{Nucl. Phys.} \textbf{B342} (1990) 246.
%
\bibitem{SVV-epl}
V. Slesar, M. Visinescu and G.E. V\^ilcu,
\textit{EPL} \textbf{110} (2015) 31001.
%
\bibitem{M-S-Y}  D. Martelli, J. Sparks and S.-T. Yau, 
\textit{Comm. Math. Phys.} \textbf{268} (2006) 39.
%
\end{thebibliography}
\end{document}